\documentclass[cits]{PoS}

\title{Core-shift and spectral analysis of the 2006 radio flare in CTA102}

\ShortTitle{Core-shift and spectral analysis of the 2006 radio flare in CTA102}

\author{\speaker{Christian M. Fromm}$^a$ \thanks{Member of the Max Planck Research School (IMPRS) for Astronomy and Astrophysics at the Universities of Bonn and Cologne.}, Eduardo Ros$^{b,a}$, Manel Perucho$^b$,  Tuomas Savolainen$^a$,\newline Petar Mimica$^b$, Andrei P. Lobanov$^a$ and J. Anton Zensus$^a$\\
\llap{$^a$} Max Planck Institut f\"ur Radioastronomie, Auf dem H\"ugel 69, 53121 Bonn, Germany\\
\llap{$^b$} Departament d'Astronomia i Astrof\'\i sica, Universitat de Val\`encia, Dr. Moliner 50, E-46100 Burjassot, Val\`encia, Spain\\
       E-mail: \email{cfromm@mpifr.de}}

\abstract{The 2006 radio flare in the blazar CTA102 is the largest ever reported for this source. The analysis of the single--dish light curves revealed a possible shock-shock interaction as  a scenario for the 2006 outburst. In order to confirm this hypothesis we analyzed eight multi-frequency (2\,GHz - 86\,GHz) VLBI observations covering the 2006 flare. We detected a radio component ejected around 2005.9 and a stationary feature around 0.1\,mas away from the core. 
We also performed a core-shift and spectral analysis on the fully calibrated VLBI maps. The frequency dependent position of the core slightly deviates from the typical $\nu^{-1}$-dependence during the flaring state. This behaviour may reflect a mismatch between magnetic energy density and energy density of the relativistic particles. From the core-shift corrected maps we obtained the spectral parameters and computed the temporal and spatial evolution of the jet intrinsic parameters such as the magnetic field and the particle density. The overall picture of the source could be best interpreted as an over-pressured jet perturbed by traveling shock waves.}

\FullConference{11th European VLBI Network Symposium \& Users Meeting\\
		 9-12 October 2012\\
		 Bordeaux (France)}

\begin{document}

\section{Introduction}
The technique of Very Long Baseline Interferometry (VLBI) offers the unique capability to resolve the structure of radio jets in Active Galactic Nuclei (AGN). Hence, we can study the morphology of those jets and trace the kinematical variations within long--term monitoring programs such as MOJAVE \cite{lis09}. Extending the single--frequency VLBI observations to multi--frequency ones provides additional access to the spectral evolution in AGN jets from which the magnetic field and particle densities are derived \cite{lob98a,sav08}. The conditions in the direct vicinity of the black hole can be probed by analysing the frequency--dependent position of the innermost region of the jet, the so-called core-shift \cite{lob98b,hir05}.

The dense sampling of the 2006 radio flare in the active galaxy CTA\,102 ($z$=1.037) with multi--frequency single--dish as well as  VLBI observations provides an excellent laboratory to study the temporal and the spatial variation in the physical conditions  of the source during the quiet and active states. The analysis and modelling of the single dish light curves revealed a travelling shock--recollimation shock interaction in a non-pressure matched jet as a possible explanation of the observed behaviour \cite{fro11}. Additional evidences supporting this hypothesis were found from the kinematic analysis of the long-term monitoring of the source at 43\,GHz within the Boston University Blazar Monitoring Program \cite{jor05}, at 15\,GHz with the MOJAVE program \cite{lis09}, and from eight multi-frequency VLBI observations (2\,GHz - 86\,GHz) during the 2006 outburst. This study detected a new traveling component, which was ejected in 2005.9 with an apparent speed of (13$\pm$2)\,c. Furthermore, several stationary features along the jet were identified with the first one located at 0.1\,mas from the radio core. The interaction between the travelling and the first stationary component occurred around 2006.3, which corresponds to the peak in the 37\,GHz single--dish light curve and supports the hypothesis of a shock--shock interaction \cite{fro12}.

Here, we present a core-shift and a spectral analysis on the fully calibrated multi-frequency VLBI observations of CTA\,102 to further investigate the possible interaction between a travelling shock and a recollimation shock. Throughout the paper we define the spectral index, $\alpha$, using the relation $S_\nu \propto \nu^{\alpha}$. The optically thin spectral index, $\alpha_0$, can be derived from the spectral slope, $s$ of the relativistic electron distribution ($N\propto E^{-s}$), via the relation $\alpha_{0}=-(s-1)/2$. We define the optically thin spectral index as $\alpha_0<0$.

\section{Data and Data analysis}
\label{data}
\begin{figure*}[t!]
\centering 
\includegraphics[width=13cm]{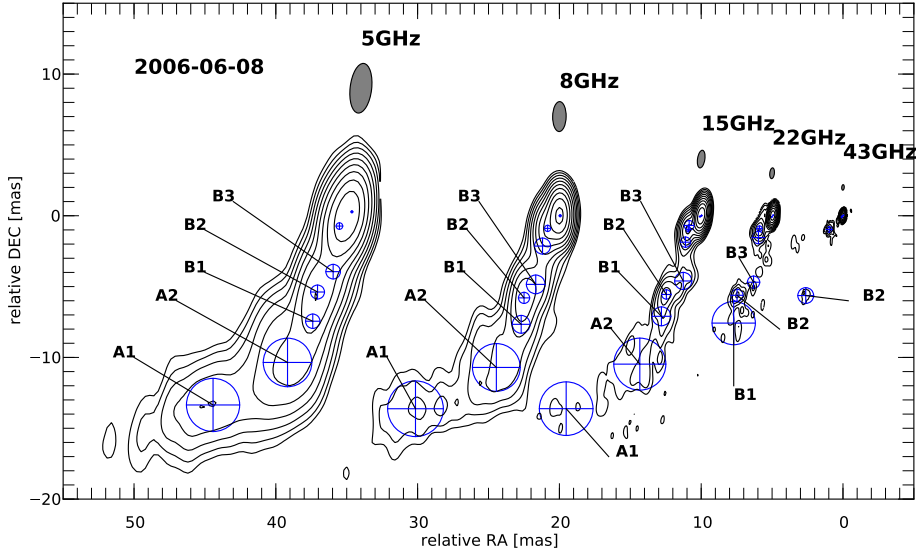} 
\caption{Uniform weighted VLBA CLEAN images with fitted circular Gaussian components at different frequencies for the July 2006 observation of CTA102. The lowest contour is plotted at $10\times$ the off-source $\mathrm{rms}$ at $43\,\mathrm{GHz}$ and increases in steps of 2. The observing frequency and the restoring beam size are plotted above each map. For the labeling we use capital letters for the same physical region in the jet and the numbers increase with inverse distance from the core. We adopt the following cosmological parameters: $\Omega_m=0.27$, $\Omega_\Lambda=0.73$ and $H_0=71\,\mathrm{km\,s^{-1}\, Mpc^{-1}}$. This results in a linear scale of $8.11\,\mathrm{pc\,mas^{-1}}$ or $26.45\,\mathrm{ly\,mas^{-1}}$ for CTA\,102 ($z$=1.037). With these conventions, $1\,\mathrm{mas}\,\mathrm{yr}^{-1}$ corresponds to $52.9\,\mathrm{c}$. } 
\label{allcont} 
\end{figure*}

We used eight multi-frequency VLBI observations from May 2005 until April 2007 spanning a frequency range from 2\,GHz to 86\,GHz for our analysis. We used AIPS for the data calibration and DIFMAP for the imaging and model fitting (see \cite{fro12} for further details). Figure \ref{allcont} shows the uniform weighted VLBA clean images with fitted circular Gaussian components for the June 2006 observations at different frequencies. Since the absolute position of the source is lost during the data calibration, we used a 2D cross-correlation technique to reconstruct the relative position among different frequency images taking into account the uneven (u,v) coverage between them. If we assume that the extended jet structure is optically thin, e.g., its location is frequency independent, we can use this structure to correct for the opacity shift of the innermost regions \cite{fro13,cro08}. From the obtained shift and the position of the core at different frequencies we calculated the core-shift and derived the magnetic field and particle density following \cite{lob98b,hir05}. Once the images were corrected for the opacity shift, we convolved them with a common set of image parameters (i.e., beam and pixel size), stacked them together and fitted a homogeneous synchrotron self-absorbed spectrum to each pixel to obtain the turnover frequency, $\nu_m$, the turnover flux density, $S_m$, and the optically thin spectral index, $\alpha_0$. A homogenous synchrotron self-absorbed spectrum is given by
$S_\nu \approx S_m\left({\nu}/{\nu_m}\right)^{5/2}\left(1-\exp{\left(-\tau_m\left(\nu/\nu_m\right)^{\alpha_0-5/2}\right)}\right)/\left(1-\exp{(-\tau_m)}\right)$,
where $\tau_m\approx3/2\left(\left(1-{4\alpha_0}/{15}\right)^{1/2}-1\right)$ is the optical depth at the turnover. The uncertainties of the extracted turnover values were computed from Monte-Carlo simulations taking into account the uncertainties of the image alignment and the SNR--based flux density error in each pixel. 

\section{Core-shift  and spectral analysis}
\label{specana}

\begin{figure}[h!]
\resizebox{\hsize}{!}{\includegraphics{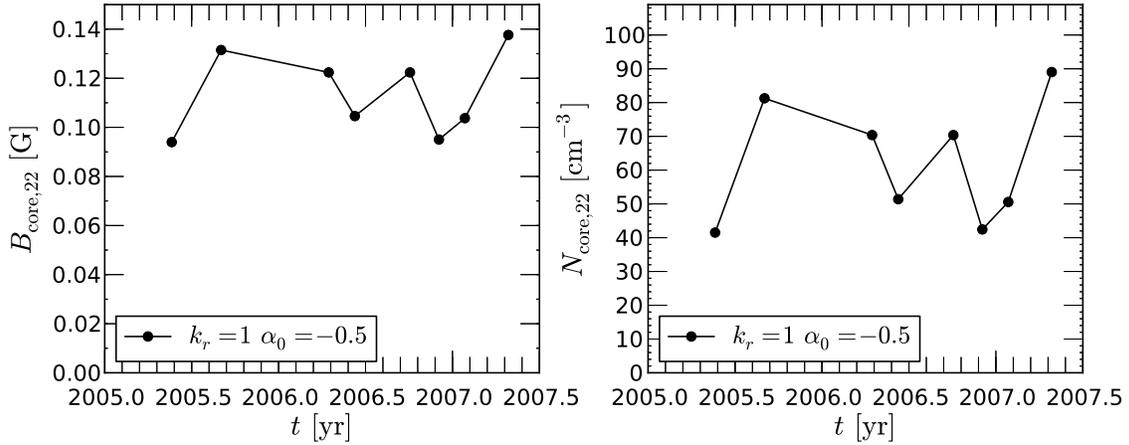}} 
\caption{Evolution of the core magnetic field, $B_\mathrm{core}$ (top), and the relativistic particle density at the core, $N_\mathrm{core}$ (bottom), as derived from the core--shift analysis.}
\label{csplot} 
\end{figure}

The frequency--dependent core position can be approximated by $r_\mathrm{core}=A\left(\nu^{-1/k_r}-\nu_\mathrm{ref}^{-1/k_r}\right)$,
where $\nu_\mathrm{ref}$ is the reference frequency, and $k_r$ reflects the physical conditions in the source, i.e., the evolution of the particle density and the magnetic field along the jet\footnote{$k_r=((3-2\alpha_0)b+2n-2)/(5-2\alpha_0)$ where $b$ parametrizes the evolution of the magnetic field $B\propto R^{-b}$ and $n$ the evolution of the particle density $N\propto R^{-n}$. If there is equipartition between magnetic energy density and energy density of relativistic particle, $k_r=1$ for $b=1$ and $n=2$.} \cite{lob98b,hir05}. We applied this approximation to the measured core-shfits and obtained an average value of  $\langle k_r\rangle=0.8\pm0.2$ and $\langle A \rangle=(3.4\pm1.6)\,\mathrm{mas/GHz}$. Using the Equations presented in \cite{lob98b,hir05} and assuming $k_r\simeq1$, $\alpha_0=-0.5$, and a lower electron Lorentz factor at core of $\gamma_\mathrm{min,core}=100$, we computed the average magnetic field and particle density at the $22\,\mathrm{GHz}$ core to be $\langle B_\mathrm{core,22\,\mathrm{GHz}}\rangle= 100\,\mathrm{mG}$ and $\langle N_\mathrm{core,22\,\mathrm{GHz}} \rangle=60\,\mathrm{cm}^{-3}$ (see \cite{fro13} for more details). In Fig. \ref{csplot} we present the temporal variation of $B_\mathrm{core,22\,\mathrm{GHz}}$ and $N_\mathrm{core,22\,\mathrm{GHz}}$ derived with the assumptions mentioned above. 

\begin{figure*}[h!]
\centering 
\includegraphics[width=13cm]{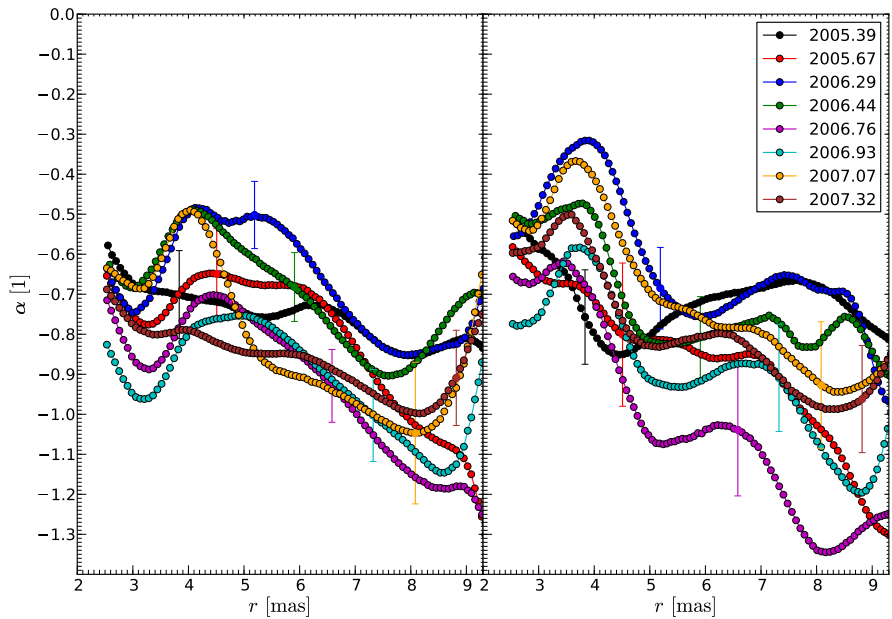} 
\caption{Evolution of the spectral index, $\alpha$  $(S_\nu\propto \nu^{\alpha})$ for $(2\,\mathrm{mas}<r<10\,\mathrm{mas})$  along the jet ridge line (see \cite{fro13b} for details on the jet ridge line). For reasons of readability only one error bar per epoch is shown. Left: values along the jet ridge line. Right: average values transversal to the jet ridge line.} 
\label{alpha} 
\end{figure*}

For a distance $r<2\,\mathrm{mas}$ from the core, the spectral turnover was within our frequency range and we were able to extract $\nu_m$, $S_m$, and $\alpha_0$. However, farther downstream, the turnover frequency moved out of our frequency range and we could only extract the spectral index, $\alpha$ ($S_\nu\propto \nu^\alpha$). For $r<2\,\mathrm{mas}$, the turnover frequency decreased from $\nu_m\sim40\,\mathrm{GHz}$ to $\nu\sim5\,\mathrm{GHz}$ with a local maximum of $\nu\sim8\,\mathrm{GHz}$ at $r=1.5\,\mathrm{mas}$. A similar behaviour was found for the turnover flux density, declining from $S_m\sim2.2\,\mathrm{Jy}$ to $S_m\sim0.2\,\mathrm{Jy}$ with a second peak at $r=1.5\,\mathrm{mas}$ of $S_m=0.6\,\mathrm{Jy}$. At larger distances from the core we found additional local maxima in the evolution of $\alpha$ at $r\sim4\,\mathrm{mas}$, $r\sim9\,\mathrm{mas}$, and $r\sim16\,\mathrm{mas}$. In Figure \ref{alpha} we show the temporal evolution of $\alpha$ along the jet for $2\,\mathrm{mas}<r<10\,\mathrm{mas}$.

Based on theory of synchrotron self-absorption we used the detected spectral turnover for $r<2\,\mathrm{mas}$ and derived the magnetic field and particle density along the jet \cite{fro13,sav08}. The magnetic field decrease from $B\sim60\,\mathrm{mG}$ at $r=0.1\mathrm{mas}$ to $B\sim2\,\mathrm{mG}$ at $r=2.0\,\mathrm{mas}$. For $r<0.6\,\mathrm{mas}$ the magnetic field geometry was best described by a toroidal field ($B\propto R^{-0.9}$) and farther downstream by a poloidal one ($B\propto R^{-2}$).  For the calculation of the particle density we assumed both adiabatic and radiative losses and computed the evolution of the upper and lower electron Lorentz factor along the jet using the derived evolution of the magnetic field. The particle density took values between $100\,\mathrm{cm^{-3}}$ and $0.8\,\mathrm{cm^{-3}}$ with a secondary maximum of $5\,\mathrm{cm^{-3}}$ at $r=1.5\,\mathrm{mas}$. Based on the magnetic field and the particle density we derived the magnetization along the jet $\sigma_\mathrm{mag}=U_b/U_e$, with $U_b=B^2/(8\pi)$ the magnetic energy density and $U_e$ the energy density of the relativistic particles. The magnetization decreased from  $\sigma_\mathrm{mag}\sim0.1$ at $r=0.1\,\mathrm{mas}$ to $\sigma_\mathrm{mag}\sim0.001$ at $r=2.0\,\mathrm{mas}$. At $r=1.5\,\mathrm{mas}$ we computed a second peak in the magnetization of $\sigma_\mathrm{mag}\sim0.03$.

\section{Discussion and Conclusions}
\label{disc}

The flaring activity in blazars is in general explained with the propagation of a relativistic shock wave through an underlying steady state jet.  At the shock front, the magnetic field and particle density increase, which leads to an enhancement of the emission (see, e.g., \cite{mar85}). Due to the limited resolution, we first detect this effect in the increased core flux density. 
From the analysis of the single--dish light curves we derived that the 2006 radio flare started around 2005.6 \cite{fro11}, which corresponds to the first increase in $B_\mathrm{core}$ and $N_\mathrm{core}$. In addition, we detected a traveling component, associated with the traveling shock around 2006.3 \cite{fro12}, which is in agreement with the decrease in $B_\mathrm{core}$ and $N_\mathrm{core}$. Thus, we interpret the variation in the magnetic field and the particle density to be due the propagation of  a travelling shock through the core region. The second increase in $B_\mathrm{core}$ and $N_\mathrm{core}$  can be explained with a new traveling shock entering the core region (see Fig. 17 in \cite{fro12}). The increase in the spectral index, $\alpha$, along the jet can be interpreted by i) traveling shocks or ii) recollimation shocks. If the increase is caused by a moving shock, the position of the peak in $\alpha$ propagates downstream and is stationary in the case of a recollimation shock. In the latter case the increase in $\alpha$ is caused by an adiabatic compression of the plasma at position of the recollimation shock \cite{mim09}. Based on this differentiation and on the results of the kinematic analysis \cite{fro12}, we assign the increase of the spectral index at $r\sim4\,\mathrm{mas}$ and $r\sim16\,\mathrm{mas}$ to recollimation shocks. Since the increase in $\alpha$ around $r\sim9\,\mathrm{mas}$ is slightly traveling downstream it could be caused by a traveling shock wave. The increase in the turnover values at $r\sim1.5$ could be the result of an interaction between a traveling shock ($\beta_\mathrm{app}\sim8.5\,c$) with a recollimation shock at $r\sim1.5\,\mathrm{mas}$ (see also Fig. 8 in \cite{fro12}). The temporal increase in the spectral index for $r<0.5\,\mathrm{mas}$ indicates the passage of a travelling component through this region. However, the limited resolution does not allow further investigation within that region.  

The derived magnetic field and particle densities from the spectral analysis are in good agreement with the ones calculated from the core-shift analysis. The obtained value for the magnetization of $\sigma_\mathrm{mag}<1$ reflects the slight departure from $k_r\neq1$ and points towards a particle--dominated jet. The change in the orientation of the magnetic field from toroidal to poloidal could reflect a spine-sheath jet, i.e., for $r<0.5\,\mathrm{mas}$ we observe mainly emission from the spine and for larger distances emission from outer sheath.

Summarising, we derived the magnetic field and particle density of the jet in CTA\,102 during the 2006 radio flare and the obtained variation is in agreement with the passage of a relativistic shock through the core region. Based on the combination of the spectral analysis presented here, and the results of the kinematic analysis in \cite{fro12} we claim that the observed behaviour in CTA\,102 is the result of the propagation and interaction of traveling shock waves with recollimation shocks in an over-pressured jet. Further analysis of the interaction between these features will be conducted using relativistic hydrodynamic (RHD) and emission simulations following \cite{per10,mim09}.

\end{document}